\documentclass{PoS}

\title{One-thimble regularisation of lattice field theories: is it only a
dream?}

\ShortTitle{One-thimble regularisation of lattice field theories}

\author{\speaker{Francesco Di Renzo}, Simran Singh and Kevin Zambello\\
        Dipartimento di Scienze Matematiche, Fisiche e Informatiche,
        Universit\`a di Parma and INFN,
Gruppo Collegato di Parma, I-43124 Parma, Italy\\
        E-mail: \email{francesco.direnzo@unipr.it},
        \email{simran.singh@unipr.it}, \email{kevin.zambello@studenti.unipr.it}}

\abstract{Lefschetz thimbles regularisation of (lattice) field theories was put forward as a possible solution
to the sign problem. Despite elegant and conceptually simple, it has many subtleties, a major one
boiling down to a plain question: how many thimbles should we take into account? In the original
formulation, a single thimble dominance hypothesis was put forward: in the thermodynamic limit,
universality arguments could support a scenario in which the dominant thimble (associated to the
global minimum of the action) captures the physical content of the field theory. We know by now
many counterexamples and we have been pursuing multi-thimble simulations ourselves. Still, a
single thimble regularisation would be the real breakthrough. We report on ongoing work aiming
at a single thimble formulation of lattice field theories, in
particular putting forward the proposal of performing Taylor expansions
on the dominant thimble.}

\FullConference{37th International Symposium on Lattice Field Theory - Lattice2019\\
		16-22 June 2019\\
		Wuhan, China}

\begin{document}

\section{Thimbles and single thimble dominance hypothesis in a nutshell}

The sign problem is a major obstacle to lattice simulations of
theories we would be interested in, among which QCD at finite
baryon density. The problem is quite easily described\footnote{In the following
we adopt a light notation in which a field theory looks like an
ordinary integral.}: we want to compute
$$
< O > \, = \, \frac{1}{Z} \, \int dx \; e^{-S(x)} \, O(x) \;\;\;\;\;\;\;\;
\mbox{with} \;\; S(x) = S_R(x) + iS_I(x)
$$ 
but (with a complex action in place) $e^{-S}$ can not be regarded as a decent
(positive) probability measure and Monte Carlo simulations are not
viable. Thimble regularisation \cite{Aurora,Kikukawa} is easily described as well
\begin{itemize}
\item One complexifies the degrees of freedom, {\em i.e.} $\,x
  \rightarrow z=x+iy$ and $\,S(x) \rightarrow S(z)$.
\item One then looks for critical points $p_{\sigma}$ where $\partial_z S = 0$.
\item The thimble $\mathcal{J}_\sigma$ attached to each critical point is the union of all
  the Steepest Ascent paths (SA) which are the solutions of $\frac{d}{dt} z_i = \frac{\partial
  \bar{S}}{\partial \bar{z_i}}$ stemming from the critical point
(initial condition). 
\item Due to the olomorphic nature of $S$, the original
  integral is convergent on the thimble, with $S_I$ staying constant
  (so that the sign problem is killed).
\end{itemize}
Thimbles are manifolds of the same (real) dimension of the
original manifold the theory was formulated on, but they are embedded in
a manifold of twice that dimension. The integration measure on the
thimble encodes the orientation of the latter with respect to the
embedding manifold and this sadly reintroduces a {\em residual} sign
problem due the so-called residual phase. We have no simple recipe to
compute the integration measure other than the following one, which is
easy to understand, but heavy to implement
\begin{itemize}
\item At the critical point one has to solve the Takagi problem for
  the Hessian of the action $H(S,p_{\sigma})\,v^{(i)} = \lambda_i \,
  \bar{v}^{(i)}$.
\item The Takagi values $\lambda_i$ fixe the rate at which the real
  part of the action increases along the SA paths.
\item The Takagi vectors $v^{(i)}$ provide a basis for the tangent
  space at the critical point.
\item The tangent space at each point on the thimble can be
  reconstructed by parallel transporting the Takagi vectors along the
  SA paths.
\end{itemize}
All in all, Lefschetz/Picard theory states that a thimble decomposition for the original integral holds
\begin{equation}
<O> \, = \, \frac{\sum_{\sigma} \; n_{\sigma} \,
  e^{-i\,S_I(p_{\sigma})} \, \int_{\mathcal{J}_\sigma} dz \;
  e^{-S_R}\; O\; e^{i\,\omega}}{\sum_{\sigma} \; n_{\sigma} \, e^{-i\,S_I(p_{\sigma})}\, \int_{\mathcal{J}_\sigma} dz \;
  e^{-S_R}\; e^{i\,\omega}}
\label{eq:ThimbleDecomposition}
\end{equation}
In (\ref{eq:ThimbleDecomposition}) the harmless constant phase factors due to
$S_I$ are factored in front of the integrals, while the residual sign
problem is due to the residual phases $e^{i\,\omega}$. Both the numerator
and the denominator ({\em i.e.} the partition function) receive
contributions in principle by all the critical points, even if the {\em
intersection numbers} $n_{\sigma}$ can be zero for possibly many
critical points. Actually $n_{\sigma}=0$ for a critical point when the
associated {\em instable} thimble does not intersect the original
integration manifold\footnote{The instable thimble is the union of
the Steepest Descent (SD) paths stemming from a critical point.}.\\

Collecting contributions from multiple thimbles can indeed
make thimble regularisation a hard problem. Not only one has to deal
with the computation of many contributions; the combination of many
terms in the numerator (and denominator) of
(\ref{eq:ThimbleDecomposition}) can actually result in a renewed sign
problem, given the multiple phase factors
$e^{-i\,S_I(p_{\sigma})}$. In the original proposal a {\em single
  thimble dominance} hypothesis was put forward. There could be
situations in which the dominant thimble alone ({\em i.e.} the one
associated to the absolute minimum of the real action) could encode
the result one is interested in. A first argument is a very simple
one: from semiclassical arguments, the contribution from the global
minimum of the action is more and more enhanced in the thermodynamic
limit. Moreover, universality arguments can be taken into account: 
the dominant thimble regularisation defines a local QFT with exactly 
the same symmetries, the same number of degrees of freedom (belonging 
to the same representations of the symmetry groups) and the same local 
interactions as the original theory. Moreover, the perturbative
expansion is the same as in the original formulation. While these
arguments are not enough to draw a definite conclusion, it was
reassuring that in the first application of thimble regularisation
(the relativistic Bose gas) the approximation proved to work very
well \cite{BoseGAS}.\\
However, it did not take that long for counterexamples to show up. In
the case of the Thirring model it was shown that the dominant thimble
could not capture the (complete) correct result
\cite{ThirringKiku,PauloAndrei1}. This was one of the motivations for
exploring alternative formulations inspired by thimbles. The idea of
complexifing the degrees of freedom (in general, of deforming the
original domain of integration) is in fact a very general
one. Alternatives to thimbles appeared, {\em e.g.} the holomorfic
flow \cite{PauloAndrei1} or various approached to the query for
sign-optimised manifods, possibly enforced by deep-learning techniques 
\cite{PauloAndrei2,PauloAndrei3,Mori}

\subsection{Multiple thimbles simulations}

Defining
\begin{equation}
<O>_{\sigma} \, = \, \frac{\int_{\mathcal{J}_\sigma} dz \;
  e^{-S_R}\; O}{\int_{\mathcal{J}_\sigma} dz \;
  e^{-S_R}} \, = \, \frac{\int_{\mathcal{J}_\sigma} dz \;
  e^{-S_R}\; O}{Z_{\sigma}}
\label{eq:SingleThimbleVEV}
\end{equation}
we can rewrite (\ref{eq:ThimbleDecomposition}) as
\begin{equation}
<O> \, = \, \frac{\sum_{\sigma} \; n_{\sigma} \,
  e^{-i\,S_I(p_{\sigma})} \, Z_{\sigma} <O\,e^{i\,\omega}>_{\sigma} }{\sum_{\sigma} \; n_{\sigma} \, e^{-i\,S_I(p_{\sigma})}\, \, Z_{\sigma} <e^{i\,\omega}>_{\sigma}}.
\label{eq:ThimbleDecompositionBis}
\end{equation}
(\ref{eq:SingleThimbleVEV}) has the obvious interpretation of an
expectation value ({\em VEV}) on a {\em single} thimble, with the measure given
by the real part of the action. (\ref{eq:ThimbleDecompositionBis}) can
in turns be intepreted as a {\em weighted sum of VEV contributions}, the
weights being given by coefficients involving the $Z_{\sigma}$. Stated
in this way, the task of performing multiple thimbles simulations
can be reduced to {\em (a)} performing single thimble simulations;
{\em (b)} computing the relative weights in
(\ref{eq:ThimbleDecompositionBis}). As one could expect, it turns out
that {\em (b)} is a harder task than {\em (a)}. Nevertheless there were
cases in which we could perform multiple thimbles simulations,
according to two different strategies.
\begin{itemize} 
\item There are cases in which it turns out that only a limited number of
  thimbles contributes, maybe also in presence of symmetries ensuring
  that a few contributions are equal. This turned out to be the case
  for QCD in $0+1$ dimensions \cite{QCD01}, in which a successful computation came
  out of the sum of only two contributions (this is just a case in
  which a symmetry is in place) according to
\begin{equation}
<O> \, = \, \frac{<O\,e^{i\,\omega}>_{\sigma_1} \,+\, \alpha \, <O\,e^{i\,\omega}>_{\sigma_2}}{<e^{i\,\omega}>_{\sigma_1} \,+\, \alpha \, <e^{i\,\omega}>_{\sigma_2}}.
\label{eq:ThimbleDecompAlpha}
\end{equation}
Here the idea is to rewrite (\ref{eq:ThimbleDecompositionBis}) putting
all our ignorance into a single parameter and giving up hope of a
first principles derivation of relative weights. One should instead
fix the value of $\alpha$ assuming one known measurement as a 
normalization point and then predicting the value of other
observables. Something similar can be put at work in the playground
that first revealed the failure of single thimble simulations, {\em
  i.e.} the Thirring model \cite{Kevin}.
\item Relative weights can in turns be computed in a semiclassical
  approximation (according to what is also referred to as the {\em
    gaussian approximation}). A possible strategy is to start with
  this semiclassical computation and then compute corrections to it. It turned out
  that this works pretty well in the context of a minimal version of 
the so-called Heavy Dense approximation for QCD \cite{HDqcd}.
\end{itemize} 
All in all, we were able to show that some steps can be taken in the
direction of multiple thimbles simulations. The path to success is
nevertheless a difficult one and in the end the real breakthrough
would be going back to the idea of (some form of) one thimble simulation. 

\section{Thimble decomposition and Stokes phenomena}

In order to gain some understanding on the thimble decomposition, one
should consider the situations in which it
fails. This is when a Stokes phenomenon occurs. A Stokes phenomenon
takes place when two different critical points are connected by a SA/SD
path. This simply means that the stable thimble of one critical point
sits on top of the unstable thimble of another one. Under this
conditions there is no thimble decomposition. It turns out that
changes in the thimble decomposition can be traced back to the occurence
of Stokes phenomena: a very effective
description of all this can be found in \cite{StudyThirringKiku} (just in
the case of the Thirring model). A semplified, intuitive picture of
the relationship between thimble decomposition and Stokes phenomena
can be given as in the following.
\begin{itemize}
\item A thimble decomposition is in place when the union of a given
  number of thimbles is essentially a deformation of the original
  integration contour, just like in applications of Cauchy theorem
  (this was just the spirit of \cite{PauloAndrei1}).
\item As they are solutions of the same differential equation subject
  to different initial conditions, different thimbles can not cross
  each other. This in turns means that they act as barriers to each
  other in the thimble decomposition: when the union of a given number
  of thimbles is a {\em correct} deformation of the original 
integration contour, other thimbles are simply kept out.
\item Moving around in the parameter space of a given theory, thimbles
  do move around in the manifold embedding the original one, but they
  do it smoothly, {\em i.e.} they are always subject to the constraint of
  not crossing each other. Thus the thimbles that contribute to the
  decomposition of the original domain of integration keep on keeping
  the others out.
\item There is only one way thimbles can cross each other and this is
  just when two thimbles sit on top of each other (two different
  critical points are connected by a SA/SD path and the stable thimble
  of one sits on top of the unstable thimble of the other). When this
  occurs we are in presence of a Stokes phenomenon and thimble
  decomposition fails.
\item After a Stokes phenomenon has occured, the {\em relative}
  arrangement of thimbles can change and a different thimble
  decomposition is in place.
\end{itemize}

\section{Taylor expansions on Lefschetz thimbles}

A very important point we want to stress is that Stokes phenomena mark
discontinuities in the thimble decomposition, {\em i.e.} in the
coefficients $n_{\sigma}$. This does not mean that physical
observables are discontinuous. Continuity of physical observables
across discontinuities of the $n_{\sigma}$ coefficients ({\em i.e.} of
the thimble decomposition) is indeed a possible handle on fixing the
values of the $n_{\sigma}$ themselves\footnote{See the discussion of the
simple $\phi^4$ toy model in \cite{thimbleRMT}.}. This is not the end
of the story, and the continuity of physical observables across Stokes
phenomena has yet another interesting application.\\
The idea is to Taylor expand an observable around a point $\mu_0$
where one single thimble is enough to reconstruct the correct result
\begin{equation}
\langle O \rangle (\mu)= \langle O \rangle (\mu_0) + \left. \frac{\partial \langle O \rangle}{\partial \mu} \right|_{\mu_0} (\mu - \mu_0) + \frac{1}{2} \left. \frac{\partial^2 \langle O \rangle}{\partial \mu^2} \right|_{\mu_0} (\mu - 
\mu_0)^2 + \ldots 
\end{equation}

\begin{figure}[bp]
	\centering
	\includegraphics[height=5.6cm]{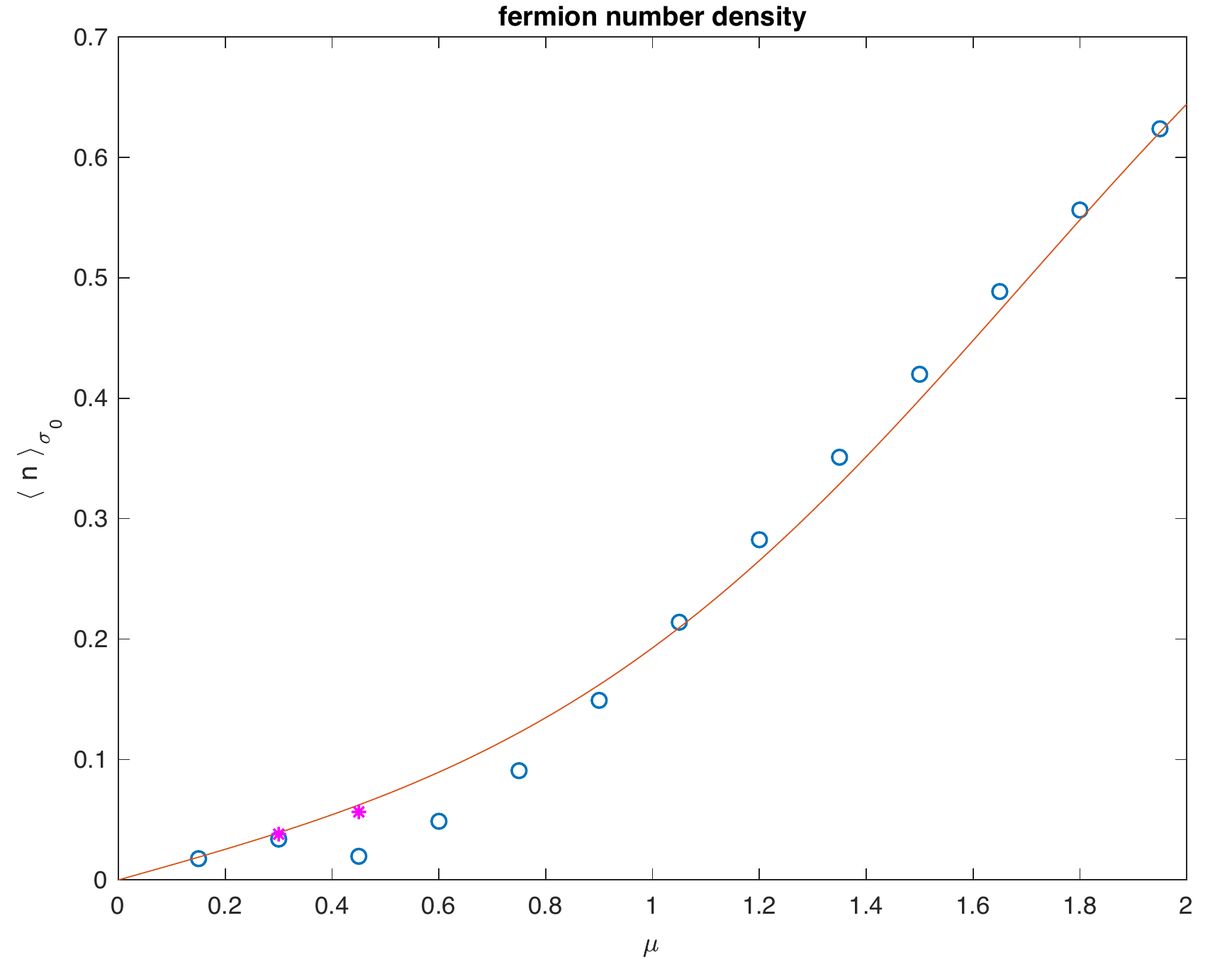}
	\caption{The fermion number density for the $L=2$, $\beta=1$, $m=1$
          Thirring model. The solid line is the (known) exact result. Circles
are the result of a single thimble computation. Stars
are the result of the Taylor expansion.}
\end{figure}

All in all, we compute Taylor expansion coefficients on a single
thimble in a region in which this is enough and then we reconstruct
values of the observable in a region in which multiple thimbles would
be needed to reconstruct the correct result. \\
A toy model application is displayed in Figure 1: this is the
computation of the fermion number density for the $L=2$ Thirring model 
($\beta=1$, $m=1$). The solid line is the (known) exact result. Circles
are the result of a single thimble computation: as one can see, in an
extended region this fails to reconstruct the correct result. Stars
are the result of the Taylor expansion ({\em i.e.} a few steps taken
into the {\em bad} region by computing Taylor coefficients in the {\em
good} one). A few more details can be found in \cite{Kevin}.\\
This is not the end of the story. After the conference we were able to
show that by bridging different regions in which one single thimble is
enough to evaluate Taylor coefficients, one can in some cases
circumvent the necessity of performing multiple thimbles simulations.  

\section{Conclusions}

The ideas that were discussed at the conference were
admittedly very prototypal ones and only a very basic example of performing
Taylor expansions on Lefschetz thimbles was provided. After the
conference we made some progress, which will be the subject of a paper
which will be issued soon. All in all, we are confident that the idea
of having Taylor expansions bridging regions in which single thimble
simulations are correct can indeed (in some cases) circumvent the
necessity of multiple thimble simulations. 

\section*{Acknowledgements}
This work has received funding from the European Union's Horizon 2020
research and innovation programme under the Marie Sk{\l}odowska-Curie
grant agreement No. 813942 (EuroPLEx). We also acknowledge support from I.N.F.N.
under the research project {\sl i.s. QCDLAT}.

\end{document}